# Temperature dependence of the superconductor energy gap

Ralph C. Dougherty,[1,*] and J. Daniel Kimel[2,†]

[1]*Department of Chemistry & Biochemistry, Florida State University, Tallahassee FL 32306-4390 USA*
[2]*Department of Physics,* Florida *State University, Tallahassee FL 32306-4350 USA*

Magnitude of the temperature dependent energy gap, the gap between the energy of the Fermi level and the next available electronic energy level in the system, for a superconductor depends strongly upon the superconductor's internal magnetic field. Because of this it is necessary to specify the internal magnetic field of a superconductor to make coherent remarks concerning its temperature dependent energy gap. For a superconductor to be able to receive an external magnetic field, there must be a vacant energy state in the superconductor to receive the energy associated with the field. For a small range of energies near that of the critical magnetic field, $H_c$, these energy states lie within the superconductor temperature dependent energy gap. This paper uses thermodynamic analysis of the energy balance in the loss of dissipative electron scattering and the change in entropy of the conducting phase that occur in the phase transition between the normal metal and the superconducting state to suggest that changes in electron Gibbs free energy at $T$, from these sources are the basis for the temperature dependent energy gap. The critical magnetic field for a superconductor at temperature, $T$, $H_c(T)$ occurs when the energy of the magnetic field is equal to the magnitude of the superconductor energy gap at $T$. Under these conditions, the internal magnetic field of the superconductor corresponds to $H_c$. When the superconductor energy gap is occupied with the energy of the external magnetic field, the normal metal conducting bands that became inaccessible at the superconductor — normal conductor phase transition are once again available for conduction, and the superconductor quenches. Origins of the superconductor energy gap arising from loss of dissipative electron scattering and development of coherent electron lattice order at the superconductor phase transition lie in the laws of thermodynamics, which cannot be casually neglected. Experimental data from the literature suggests that the ratio of the superconductor energy gap to the superconductor critical temperature depends upon the chemical structure of superconductor. We anticipate that the superconducting energy gaps for mercury, and lead will show small maxima near 0.21, and 0.11 K, respectively. Anticipated maxima for type I superconductors are due to the two sources of entropy differences between the normal phase and the superconducting phase: 1) dissipative electron scattering in the normal phase; and 2) coherent order in the superconducting phase. The dissipative scattering component of the free energy depends upon the existence of a current, which is often neglected in the study of circuits. In this case the current in the superconducting state is essential to operation of the Meissner Ochsenfeld effect. In some of the elemental superconductors for which high quality $H_c$ data is available, hypothetical curve maxima are anticipated at sub-zero temperatures. The fact basis supporting these details was presented in the late 1950's, and the original authors were careful to point out the significant differences in properties for individual elements.

## I. INTRODUCTION

Transition from normal conductor to superconductor involves an explicit change in entropy for the conductor. The recent experimental demonstration of Landauer's principle, which links thermodynamics and information by Lutz, et al[1], confirms the need for a change in system conduction free energy at the phase transition between a functioning normal metal conductor and the corresponding superconductor. We submit that this is one of the central questions that must be addressed in superconductivity. Free energy difference between dissipative and non-dissipative conduction is one of the subjects of this paper. Other subjects include: the connection between this free energy difference and the temperature dependent energy gap in a superconductor; and the contribution of the entropy difference between the bulk normal conductor, which is not a wave mechanically coherent object, and the bulk superconductor, which is a wave mechanically coherent object, to the energy gap of the superconductor. A model for the superconductor energy gap, as reflected in the magnitude of the superconductor critical magnetic field, $H_c$, is developed. Graphical evaluation of the shape of the energy gap curve with $T$, for a data series from the literature, is provided in the discussion.

Electron free energy differences between dissipative and non-dissipative current flow in metals do not appear to have received much attention in the literature, though non-dissipative superconductors have been known since 1911. Entropy differences between dissipative and non-dissipative electron currents must contribute to the temperature dependent energy gap that characterizes the non-dissipative currents in superconductivity.

This paper points out the relationship between critical magnetic field for a superconductor, $H_c(T)$, and the superconductor energy gap, $\Delta E(T)$, that is an essential feature of the superconducting state. When an external critical magnetic field has been applied to a superconductor, if the external field just matches the internal field, which arises due to the Meissner Ochsenfeld effect, the external field penetrates the superconductor and fills the superconductors energy gap with it's energy, proportional to $H_c^2$.[2] When the energy of the external magnetic field is equal to the energy gap of the superconductor, Dirac fermions in the superconducting state absorb this energy and bridge the energy gap to a normal conducting state. The Fermi level increases, to its original $T_c$ level. Electrons with basis functions that produce dissipative scattering by Fermi "contact"[3] are available for conduction, and the superconductor quenches.

## II. THEORY

Gibbs free energy, equation 1, provides the foundation for understanding the possibilities for formation of an energy gap in a superconducting system.

$$G(P,T) = H - TS \qquad (1)$$

In equation 1, *G* is the Gibbs free energy, *H* is the enthalpy, and *S* is the entropy, all at *P*, and *T*. The two terms on the right of the equation provide two potential avenues for altering the free energies of systems, like normal conductors or components like conducting electrons. If we focus on the conducting electrons, which are intimately associated with the energy gap in superconductors, their enthalpy and temperature times entropy are the controlling factors. System entropy provides the only source of free energy difference between two phases, like a superconductor and the corresponding normal conductor phase, where the difference in free energy between the two phases is intrinsically temperature dependent.

### A. Electron enthalpy

It is possible to change electron enthalpy in a conductor by altering electron interactions with other particles particularly electrons and atomic nuclei. This is what happens in chemical bond formation, and in spin pairing Mott transitions, which are known to occur in specific materials including superconductors at low temperatures[4-6]. In current theory, chemical bond formation is a temperature independent process, mediated entirely by forces due to particle fields. Bond formation includes enthalpy associated with processes including: interactions of charges, coulombic attraction and repulsion; and magnetic interactions, magnetic coupling, spin pairing, etc. It corresponds to bonding enthalpy and does not seem a proper candidate for formation of a temperature dependent energy gap in superconductors. In the Mott transition, pairs of single electron conducting states are transformed into pairs of anti-bonding and bonding states, which creates a temperature independent energy gap on the insulator side of Mott systems. In this case the energy gap corresponds to the electronic excitation energy associated with the bonding and anti-bonding pair of states. These energy gaps are known to be temperature independent.

**B. Electron associated entropy**

There are two distinct sources of the energy gap in superconductors both involve entropy differences between the normal and superconducting states of the conductor. One entropy contribution to the electron free energy in superconductors comes from the current in the normal conductor at the phase transition. Entropy arises in normal conductors in the form of resistive electron scattering, which does not occur on the superconductor side of the phase transition. Involvement of electron scattering in resistivity has been known since the late 19$^{th}$ century; however, a purely electronic theory of resistivity that is consistent with the Sommerfeld equation remains to be fully developed. Sommerfeld's equation[7] gives the relationship between thermal conductivity, , and electrical conductivity, σ, known as the Wiedemann Franz Law, in terms of the temperature and established constants, equation 2.

$$\frac{\kappa}{\sigma T} = \frac{\pi^2}{3}\left(\frac{k_B}{e}\right)^2 = 2.44 \cdot 10^{-8} W\Omega K^{-2}$$

(2)

Sommerfeld's equation was developed by exclusive use of electron gas wave functions, which indicates that the Born Oppenheimer approximation is functional and lattice wave functions and their components are not involved in the relationship. This specifically means than lattice vibronic quanta, phonons, are not a part of the relationship between electrical conductivity and thermal conductivity in metals, equation 2.

In the normal conductor/superconductor phase transition, the change in entropy at the end of dissipative electron scattering is an exquisitely subtle topic in statistical thermodynamics. The subject can be considered using Landauer's principle[8,9], which links information and thermodynamics, as well as standard thermodynamic considerations. The conservative estimate given for the Landauer limit in the measurements reported by Lutz, et al.[10] is $\ln2 \cdot kT$ per event, or in this case, per electron. This data provides a credible beginning for understanding the details of the superconductor temperature dependent energy gap, at least at on-set, $T_c$. Temperature dependence for this contribution to the energy gap is linear. That is, the contribution to the electron free energy will vary as $-T\Delta S$. As the temperature decreases the contribution per electron of dissipative electron scattering in the normal phase to electron free energy in the superconducting phase will decrease.

A second contribution to electron free energy in superconductors comes from the entropy difference between the superconducting state, which is coherent, and the normal state, which is non-coherent. At the superconductor phase transition the mechanism for thermal conductivity and heat capacity fundamentally changes. On the superconductor side of the transition all of the conducting electrons are carriers of heat capacity with direct quantum mechanical coupling to the lattice dynamics. The Born Oppenheimer approximation fails here, as does the Sommerfeld equation, 2. On the normal conductor side of the phase transition some of the electrons may not be effective carriers of thermal energy, and Sommerfeld's equation, 2, functions, particularly well at very low temperatures in normal conductor systems[11].

The shape of the *ΔE v. T* curves, see figures 2-4, tells us that the energy gap increases dramatically with decreasing temperature. This increase with decreasing temperature corresponds to the effect of both the increase in current for the internal magnetic field as $H_c$ increases, and the effect of decreasing temperature on the coherence and order in the superconducting state as compared to the normal state in which the Born Oppenheimer approximation applies and there is no coherence of the electronic and vibronic wave functions.

### C. Thermodynamic model for the superconducting energy gap

When an external magnetic field, *H*, is applied to a superconductor, it is expelled by the Meissner Ochsenfeld effect[12]. When an external magnetic field penetrates the body of a superconductor, the energy contained in the external magnetic field, proportional to $H^2$, is absorbed by the superconductor energy gap. When $H = H_c$, the applied magnetic field is sufficiently large to reduce the energy gap to zero, allowing a transition to the normal conducting state. The condition $E_H = \Delta E$ enables us to calculate $H_c(T)$. Application of the standard treatment for critical phenomena[13] leads to equation 3.

$$\Delta E = a\left(1 - \frac{T}{T_C}\right)^n f(T) \qquad (3)$$

In equation 3, *n* is a critical exponent, *a* is a proportionality constant, and Δ*E* corresponds to the temperature dependent energy gap of the superconductor. The first factor on the right of equation 3 corresponds the threshold behavior near the critical temperature, $T_c$, and $f(T)$ is a slowly varying function normalized to 1 at $T = 0$. We approximate $f(T)$ by the first order McLaurin expansion, 4.

$$f(T) = 1 + b\frac{T}{T_C} \tag{4}$$

Since the energy of the uniform magnetic field is proportional to $H^2$,[2] from 3 we obtain,

$$H_C(T) = H_C(0)\left(1 + b\frac{T}{T_C}\right)^{1/2}\left(1 - \frac{T}{T_C}\right)^{n/2} \tag{5}$$

A close approximation to equation 5 is given in equation 6.

$$H_C(T) = H_C(0)\left(1 + \frac{b}{2}\frac{T}{T_C}\right)\left(1 - \frac{T}{T_C}\right)^{n/2} \tag{6}$$

Equation 6 has the potential to deal with the variety of critical magnetic curves that are presented by elemental superconductors.

### III. DATA ANALYSIS

We have used equation 6 to model the critical magnetic fields reported for 10 elemental superconductors for which we found suitable data in the literature, see table I. Only zinc, Zn and cadmium, Cd in table I have critical temperatures below 1 K. Largely because of the lack of data for low $T_c$ superconductors, the only representatives of the *d* series superconductors in table I are niobium, Nb, and tantalum, Ta.

TABLE I. Elemental superconductor critical magnetic field *v. T* data fitting parameters to equation 6.

| Elem | $H_c(0)$ | unit | b | n | $T_c$ | $\chi^2$ | COD | #pts | ref. |
|------|----------|------|------|------|-------|---------|----------|------|------|
| Al   | 106      | G    | 2.10 | 2.02 | 1.20  | 5.38    | 0.9995   | 13   | 15   |
| Zn   | 50.2     | G    | 2.52 | 2.13 | 0.907 | 1.656   | 0.9993   | 15   | 15   |
| Ga   | 52.6     | G    | 1.43 | 1.83 | 1.11  | 34.2    | 0.993    | 26   | 15   |
| Nb   | 1710     | Oe   | 2.17 | 1.94 | 9.22  | 7639.4  | 0.9996   | 70   | 16   |
| Cd   | 28.9     | G    | 1.73 | 1.86 | 0.553 | 0.312   | 0.9997   | 15   | 15   |
| In   | 294      | Oe   | 1.71 | 2.03 | 3.41  | 0.5965  | 0.999995 | 16   | 17   |
| Sn   | 317      | Oe   | 1.75 | 2.06 | 3.73  | 1.336   | 0.999996 | 36   | 17   |
| Ta   | 640      | Oe   | 1.71 | 2.03 | 3.41  | 0.5965  | 0.999995 | 12   | 17   |
| Hg   | 405      | G    | 2.30 | 2.03 | 4.15  | 3.3569  | 0.999994 | 37   | 18   |
| Pb   | 792      | G    | 2.15 | 1.96 | 7.17  | 9.0923  | 0.999995 | 34   | 19   |

It has been known for more that 50 years that heat capacity of a superconductor, the square of the superconductors critical magnetic field and the superconductor energy gap are all related[14]. The availability of a model makes it possible to enquire about the variance of the energy gap from one element to the next. This assumes that all of the superconductors in table I have the same relationship between critical magnetic field energy, $\propto H_c^2$ and the magnitude of the superconductor energy gap.

Figure 1 presents the fitted $H_c$ v. $T$ curves for the 10 superconductors in table I. If you look carefully you will see that there are three curves in figure 1 that have maxima at temperatures above ~0 K. These elements, Nb, Hg, and Pb all have values for the parameter $b$ greater than their values for the parameter $n$. Low temperature measurements for elemental mercury, Hg, look like the best case for evaluating the validity of the model in equation 6 for the critical magnetic field of a type I superconductor. Using the experimental critical magnetic field data to determine the anticipated maximum in the curve of $H_c$ v. $T$, we obtained values of 0.21, and 0.11 K for the anticipated maxima in $H_c$ for mercury and lead respectively.

From the analytical details presented in the critical magnetic field studies of the Mapother research group, see references 16-19, it is clear that as a group they were aware of the regularities of the curve shapes in the critical magnetic field data for elemental superconductors. Historically the equation used to model the temperature dependence of the critical magnetic field is shown as 7[18].

$$H_c(T) = H_c(0)\left[1 - \frac{T^2}{T_c^2}\right]$$

(7)

Mapother's group modeled the real $H_c$ data using an equation for deviations from equation 7[18]. In their analysis mercury and lead gave positive deviations and the remainder of the elemental superconductors, for which $H_c$ data was available gave negative deviations. These two groups of elements correspond one to one to elements for which parameter *b* in table I is greater than *n* (positive deviation group) and parameter *b* is less than the corresponding *n* (negative deviation group). For the type I superconductor cases in table I for which the literature provides extensive high quality data sets the coefficient of determination, COD in table I using equation 6, is essentially at the maximum. Using equation 6, for these elements, there is no need to model the deviations.

Niobium's large, relatively recent, $H_c$ data set[16], shows significant scatter using equation 6 ((1-COD)>4·10$^{-4}$). It seems likely that this is a consequence of the nearby second order phase transition to the magnetic vortex lattice[16], which has a different magnetic description than a type I magnetic field. We do not have confidence in the use of equation 6 to fit data for type II superconductors. The curve is simply for comparison.

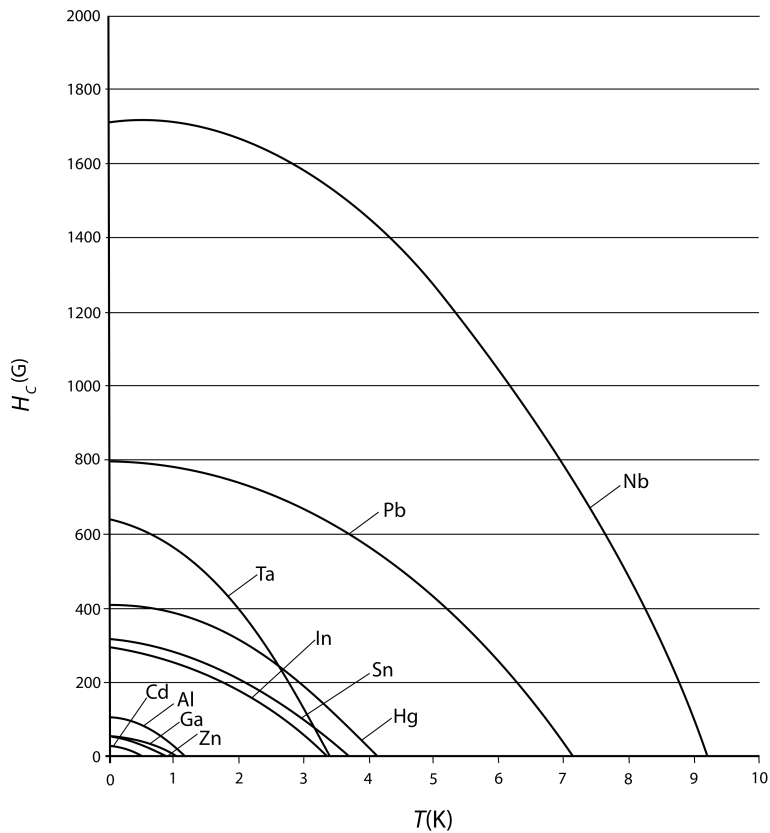

FIG. 1. Critical magnetic field curves to experimental data (equation 6) for superconductors in table I.

Lines showing the fit of the square of critical magnetic field *v. T* from equation 6 to the square of experimental data, points, for the elements listed in table I are presented in figures 2-4. Figure 2 is devoted to the *d* series elements, niobium and tantalum. Figures 3 and 4 present superconducting elements at one bar from the main group in the periodic table and column 12. Main group and column 12 elements with $T_c$ above 1.5 K are presented in figure 3 and those with $T_c$ below 1.5 K are in figure 4. The scaling factors used to reduce the scale in the three figures for the *y*-axis values are $10^4$, $10^3$ and $10^2$ for figures 2, 3, and 4 respectively.

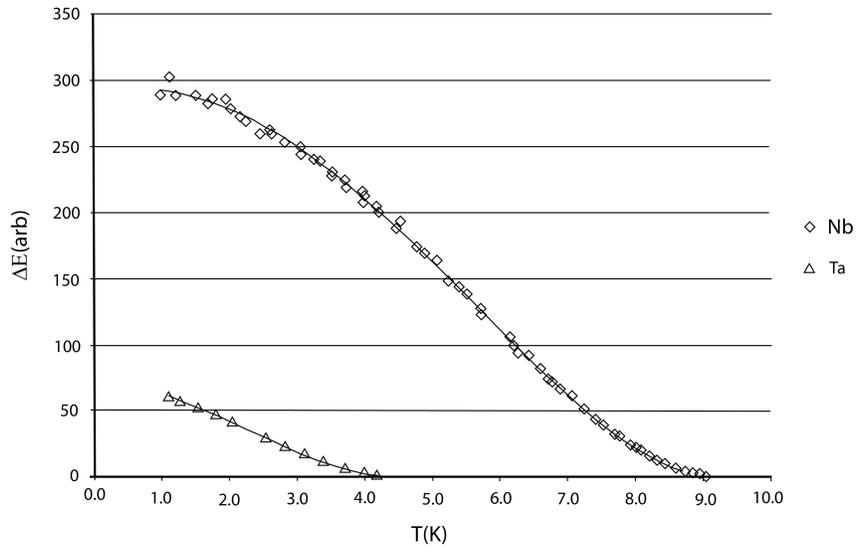

FIG. 2. Relative energy gap (arbitrary units) *v.* *T* (K) for niobium, Nb[16] and tantalum, Ta[17]. Points show the square of experimental critical magnetic field divided by $10^4$. Solid line from values obtained by equation 6, using parameters in table I on the same scale as the points.

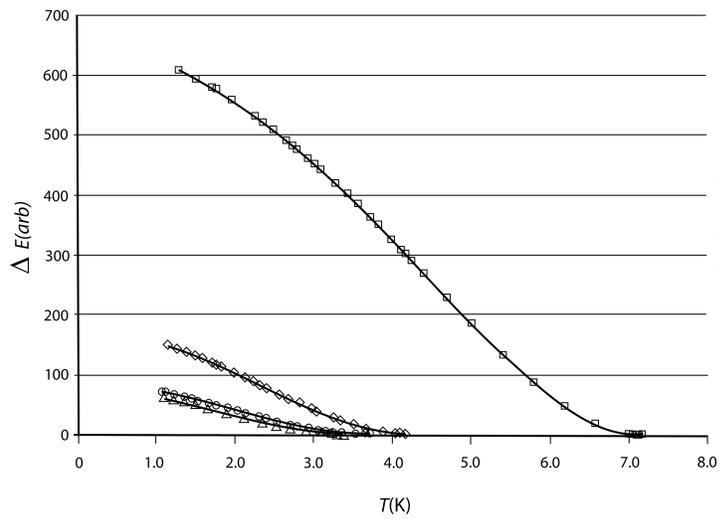

FIG. 3. Points show the square of experimental critical magnetic field divided by $10^3$. Solid line from values obtained by equation 6 using parameters in table I on the same scale as the points.

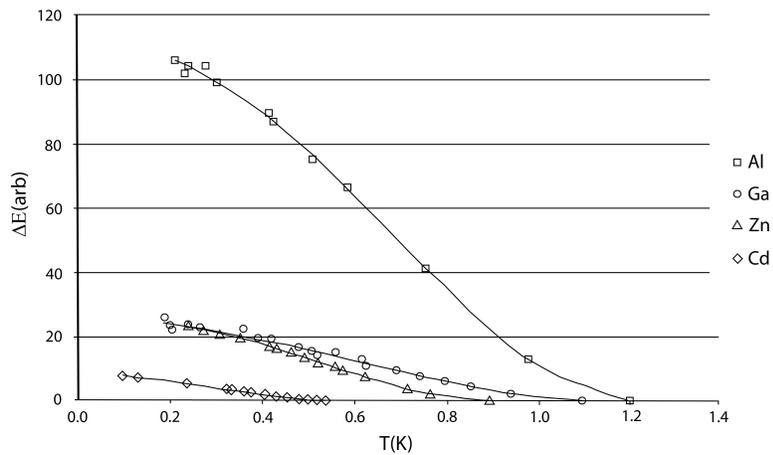

FIG. 4. Points show the square of experimental critical magnetic field divided by $10^2$ Solid line from squares of values obtained by equation 6, using parameters in table I on the same scale as the points.

## IV. DISCUSSION

Figures 2-4 illustrate the variability in the appearance of relative critical magnetic field energies and the associated superconductor energy gaps for the elements listed in table I. To illustrate the variance in the shapes of the superconductor, energy gap or critical magnetic energy, curves with temperature we have evaluated the relative energy gap, equation 8, at both 15% and 90% of the critical temperature.

$$\Delta E_{sc}(T) \simeq g H_C(T)^2 \qquad (8)$$

In equation 8, $\Delta E_{sc}(T)$ is the superconductor energy gap at the critical magnetic field as a function of temperature. For the data taken from the sources in table I, the scaling factor, $g$, is unknown, and is arbitrarily set to 1 for all relative gap comparisons for elements in table I. External magnetic fields have nothing to do with the creation of an energy gap in superconductors. The superconductor energy gap arises from the Gibbs free energy difference for the conducting electrons between the normal conductor and the superconductor. When the energy of the external magnetic field is equal to the energy gap of the superconductor, electrons in the superconducting state are able to absorb this energy and bridge the energy gap to a normal conducting state resulting in a dissipative resistance initiated quench for the superconductor. The dissipative resistance arises from the participation of electrons in basis states that support Fermi "contact" dissipative scattering in the normal conducting state. This energy gap is the maximum value for the energy gap at that temperature, which varies with the internal magnetic field of the superconductor.

The ratio of the square of the fitted values of critical magnetic field at $0.15*T_c$ to those at $0.9*T_c$ are shown in table II along with the ratio of extrapolated values (equation 6) of $\Delta E(0)$ (arbitrary units, see equation 8) to $T_c$.

There are a large enough number of elements in table II to permit the conclusion that use of a single value for energy gap at $T \approx 0$ will not be a reasonable approximation to explain the experimentally based results.

The average of the $\Delta E_{.15Tc}/\Delta E_{.9Tc}$ ration of magnetic energies in Table II is 26.2 with a standard deviation of ±4.87. This standard deviation is less than 20% of the average value for the ratios. The average value of the energy gap, estimated using critical magnetic fields for these 10 superconductors with its standard deviation was: 6.23E+4 (arb.)±9.78E+4 In this case the standard deviation is fully 1.5 times the average of the magnitude of the energy gaps divided by the respective $T_c$s.

Table II. Energy gap ratio for $0.15T_c$ to $0.9T_c$ and $\Delta E(0)/T_c$ for elements in Table I.

| Element | $\Delta E_{.15Tc}/\Delta E_{.9Tc}$ | $\Delta E(0)/T_c$ (arb.) |
|---|---|---|
| aluminum | 26.7 | 9.36E+03 |
| zinc | 32.2 | 2.78E+03 |
| gallium | 23.1 | 2.49E+03 |
| niobium | 22.1 | 3.17E+05 |
| cadmium | 23.4 | 1.51E+03 |
| indium | 19.2 | 2.53E+04 |
| tin | 33.4 | 2.69E+04 |
| tantalum | 31.5 | 1.20E+05 |

| Element | $\Delta E_{.15Tc}/\Delta E_{.9Tc}$ | $\Delta E(0)/T_c$ (arb.) |
|---|---|---|
| mercury | 25.8 | 3.94E+04 |
| lead | 22.5 | 8.77E+04 |
| average ± std dev | 26.0±4.86 | 6.32E+4±9.75E+4 |

The experimentally supported observations documented in table II, do not support a model that calls for all type I superconducting elements having the same value, e.g., 3.528 $k_B T_c$ [20], for $\Delta E(0)/T_c$ or a closely related ratio. Table II was constructed by preparing a spreadsheet using the parameters in table I, and equation 6, to model the critical magnetic fields of the 10 elemental superconductors for which we have literature data. The temperature resolution used was 0.01 K. Values from the fitted curves (equation 6) were then used with equation 8 to construct table II. For the purposes of this exercise oersteds, Oe, were considered equivalent to gauss, G.

There are five main group metals in table II. These are elements in periodic table columns 13 and 14, see table III. Values for the ratio, $\Delta E_{.15Tc}/\Delta E_{.9Tc}$, in table II for these elements range from 19.2 to 33.4 for indium and tin respectively. The wide variation in this ratio suggests chemical origins for the observed differences. These differences, like the pattern of critical temperatures for the one bar superconductors in the periodic table[21], table III, must have a chemical origin. A probable origin for this pattern is in the density of molecular electronic states in the superconductor[22].

Understanding critical temperatures in elemental superconductors, requires knowing that conducting electrons in *s* basis conduction bands will have a partial wave scattering cross section as a function of temperature down to the lowest attainable temperatures[22]. Furthermore, for *s* basis wave functions the threshold cross section is independent of temperature, so this scattering will continue at the lowest attainable temperatures. Partial wave generated electron scattering will effectively stop any phase transition for a metal to superconductivity[22]. Partial wave scattering at threshold exponentially drops with temperature as the $4l$ power for conducting electrons in wave functions with basis $l>0$[22].

TABLE III. (Color online) Periodic table showing elemental bulk superconductors at one atmosphere with their critical temperatures, $T_c$.[22, a]

| Group | 1 | 2 | 3 | 4 | 5 | 6 | 7 | 8 | 9 | 10 | 11 | 12 | 13 | 14 | 15 | 16 | 17 | 18 |
|---|---|---|---|---|---|---|---|---|---|---|---|---|---|---|---|---|---|---|
| Period 1 | H 1 | | | | | | | | | | | | | | | | | He 2 |
| Period 2 | Li 3 | Be 4 0.026 | | | | | | | | | | | B 5 | C 6 | N 7 | O 8 | F 9 | Ne 10 |
| Period 3 | Na 11 | Mg 12 | | | | | | | | | | | Al 13 1.18 | Si 14 | P 15 | S 16 | Cl 17 | Ar 18 |
| Period 4 | K 19 | Ca 20 | Sc 21 | Ti 22 0.5 | V 23 5.4 | Cr 24 | Mn 25 | Fe 26 | Co 27 | Ni 28 | Cu 29 | Zn 30 0.85 | Ga 31 1.08 | Ge 32 | As 33 | Se 34 | Br 35 | Kr 36 |
| Period 5 | Rb 37 | Sr 38 | Y 39 | Zr 40 0.6 | Nb 41 9.25 | Mo 42 0.92 | Tc 43 8.2 | Ru 44 0.5 | Rh 45 | Pd 46 | Ag 47 | Cd 48 0.57 | In 49 3.4 | Sn 50 3.7 | Sb 51 | Te 52 | I 53 | Xe 54 |
| Period 6 | Cs 55 | Ba 56 | La 57 6.0 | Hf 72 0.38 | Ta 73 4.4 | W 74 0.01 | Re 75 1.7 | Os 76 0.7 | Ir 77 0.1 | Pt 78 | Au 79 | Hg 80 4.15 | Tl 81 2.4 | Pb 82 7.2 | Bi 83 | Po 84 | At 85 | Rn 86 |

Legend: $T_c \geq 1.5$ K; $1.5 > T_c > 0.1$ K; $T_c \leq 0.1$ K. Example: Symbol Nb, Atomic Number 41, Critical Temperature ($T_c$) in Kelvin (K) = 9.25.

[a]Period 7, and the *f* elements in period 6, with the exception of lanthanum, La, are not shown.

In developing an understanding of the chemical origins of the pattern of critical temperatures for the one bar superconductors in the periodic table, we did two things: 1) used all the data for the one bar superconductors above period 7; and 2) focused on the magnitude of $T_c$ for each element.[21] The relatively high $T_c$ superconductors in the *d* series, superconductors in columns 3 through 9 in table III, are in odd numbered columns. This suggested that the Fermi level for the superconductors in columns 3, 5 and 7 of table III has a closed valence level *s* sub-shell for the metallic state of these conductors at very low temperatures. For the main group superconductors in table III, it is possible for superconductors in columns 13 and 14 to be conductors with a closed *s* sub-shell at low temperatures. This effect is classically known as the "inert pair" effect, in chemistry. Having a closed *s* sub-shell would make these conductors *p* wave superconductors at temperatures below $T_c$[23].

Mercury's relatively high $T_c$ of 4.15 K seems likely to be the result of the fact that metallic mercury can form two electron 6*s* bonds in the solid state that leave low temperature metallic mercury as an effectively closed *s* sub-shell element. Zinc and cadmium, in the same column of the periodic table as mercury, are not known to form strong *s*-*s* single bonds as are known for mercurous salts like calomel, $Hg_2Cl_2$, no "inert pair" effect.

The possibility of using the same strategy for examining the chemical basis for variance in the elemental superconductor energy gap that was used for critical temperatures is seriously limited by the relatively small amount of data that is available on critical magnetic fields and/or direct measurements of energy gaps, as compared to critical temperatures which are known for all elemental superconductors. Since the problem of the dispersion of values for the magnitude of the energy gap at near zero temperatures in superconductors is a problem on the chemical side of chemical physics, it is hoped that this problem will attract the attention of condensed matter scientists with a chemical background.

In establishment of the current paradigm for understanding superconductivity it appears that a very limited sample of superconductors were used to form the conclusion that the superconductor energy gap at ~0 K divided by the critical temperature for the superconductor was a constant for temperatures near 0 K that did not vary from one superconductor to the next[24].

## V. CONCLUSIONS

Superconductor energy gaps arise from changes in system entropy between the superconductor and the normal conductor in the phase transition. On the normal conductor side of the transition the entropy change is associated with the loss of dissipative electron scattering in the phase transition to the superconductor. This entropic contribution to the system free energy gives a term that decreases in magnitude as the temperature decreases. On the superconductor side of the phase transition the change in entropy is associated with an increase in structural coherence in the superconducting phase as temperature decreases in comparison to the normal phase which is non-coherent. Judging from the fact that the extrapolated maxima in the experimental plots of $H_c$ v. $T$ are at or near ~0 K, the increase in structural order in the superconducting phase with decreasing temperature must be significant. The term in the superconductor free energy associated with the increase in structural coherence with decreasing temperature, increases as the temperature decreases. Total free energy for the superconductor as compared to the normal conductor, the energy gap, has the potential to show a maximum above ~0 K for selected cases. Mercury may be the best elemental case for observing this maximum. For mercury the superconductor energy gap maximum should occur near ~0.21 K. Another case worthy of investigation is lead, where the maximum should occur near 0.1 K.

Electron entropy changes associated with the both the end of dissipative electron scattering, and the increase in coherent structural order in the superconducting phase at the critical temperature for a superconductor appear to be the phenomenological basis for the temperature dependent superconductor energy gap. We have examined the critical magnetic field $v.$ $T$ curves for ten elemental superconductors in the literature and used the data to construct $\Delta E(T) \propto H_c(T)^2$ $v.$ $T$ curves. These curves, figures 2-4, reflect the relative magnitude of the energy gap for the respective superconductors. When the curves are placed on the same scale for type I superconductors the ratio of $\Delta E(0)/T_c$ varied from $1.20 \cdot 10^5$ (Ta) to $1.51 \cdot 10^3$ (Cd). Cadmium had the lowest $T_c$ in the group and also the lowest value of the relative energy gap to the critical temperature, $\Delta E(0)/T_c$. The experimental data indicates that for type I elemental superconductors, the ratio of the superconductor energy gap to the superconductor critical temperature depends upon the chemical structure of superconductor.

## ACKNOWLEDGMENTS

It is a pleasure to acknowledge useful discussions with Ronald J. Clark concerning bonding in mercury. It is also a pleasure to acknowledge the helpful discussions and interactions with of many of our colleagues during the development of the ideas presented here.

[*]rdougherty@fsu.edu

[†]dankimel@comcast.net